\documentclass[11pt,5p]{elsarticle}

\usepackage{enumitem}

\setlist[itemize]{leftmargin=*,labelsep=5.8mm}
\setlist[enumerate]{leftmargin=*,labelsep=4.9mm}

\usepackage[colorlinks=true]{hyperref}
\usepackage{xurl}
\usepackage{calc}
\usepackage{longtable}
\usepackage{booktabs}
\usepackage{fancyhdr}
\usepackage{amsmath}
\usepackage{amssymb}

 \usepackage{graphicx}
 \makeatletter
 \newsavebox\pandoc@box
 \newcommand*\pandocbounded[1]{
   \sbox\pandoc@box{#1}%
   \Gscale@div\@tempa{\textheight}{\dimexpr\ht\pandoc@box+\dp\pandoc@box\relax}%
   \Gscale@div\@tempb{\linewidth}{\wd\pandoc@box}%
   \ifdim\@tempb\p@<\@tempa\p@\let\@tempa\@tempb\fi
   \ifdim\@tempa\p@<\p@\scalebox{\@tempa}{\usebox\pandoc@box}%
   \else\usebox{\pandoc@box}%
   \fi%
 }
\def\fps@figure{tbp}
\makeatother

\fancypagestyle{firstpage}{
  \fancyhf{} 
  \fancyhead[L]{\includegraphics[height=3cm]{./assets/elsevier-non-solus-new-with-wordmark.pdf}}
  \fancyhead[C]{submitted to Ecological Modelling}
  \fancyhead[R]{\includegraphics[height=3cm]{./assets/X03043800.jpg}}
}

\usepackage{tikz}

\NewDocumentCommand\citeproctext{}{}

\makeatletter
\let\@cite@ofmt\@firstofone
\def\@biblabel#1{}
\def\@cite#1#2{{#1\if@tempswa , #2\fi}}
\makeatother
\newlength{\cslhangindent}
\newlength{\csllabelwidth}
\newenvironment{CSLReferences}[2] 
{\begin{list}{}{%
	\setlength{\itemindent}{0pt}
	\setlength{\leftmargin}{0pt}
	\setlength{\parsep}{0pt}
	\ifodd #1
	\setlength{\leftmargin}{\cslhangindent}
	\setlength{\itemindent}{-1\cslhangindent}
	\fi
	\setlength{\itemsep}{#2\baselineskip}}}
{\end{list}}

\journal{Ecological Modelling}
\doi{10.1016/j.ecolmodel.2024.110890}

\begin{document}

\begin{frontmatter}
\thispagestyle{firstpage}


\title{Good Modelling Software Practices}


\author%
[1]%
{%
\href{https://orcid.org/0000-0003-3483-6036}{\includegraphics[width=1em]{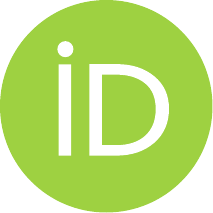}}\,
Carsten Lemmen\corref{carsten.lemmen@hereon.de}}
\author%
[2]%
{%
\href{https://orcid.org/0000-0001-6171-7716}{\includegraphics[width=1em]{logo-orcid-eps-converted-to.pdf}}\,
Philipp Sebastian Sommer}

\affiliation[1]{Institute of Coastal Systems - Analysis and Modeling,
Helmholtz-Zentrum Hereon, Max-Planck-Str. 1, 21502 Geesthacht, Germany}
\affiliation[2]{Institute of Carbon Cycles, Helmholtz Coastal Data
Center, Helmholtz-Zentrum Hereon, Max-Planck-Str. 1, 21502 Geesthacht,
Germany}

\begin{abstract}
Frequently in socio-environmental sciences, models are used as tools to
represent, understand, project and predict the behaviour of these
complex systems. Along the modelling chain, Good Modelling Practices
have been evolving that ensure---amongst others---that models are
transparent and their results replicable. Whenever such models are
represented in software, Good Modelling meet Good Software Practices,
such as a tractable development workflow, good code, collaborative
development and governance, continuous integration and deployment; and
they meet Good Scientific Practices, such as attribution of copyrights
and acknowledgement of intellectual property, publication of a software
paper and archiving. Too often in existing socio-environmental model
software, these practices have been regarded as an add-on to be
considered at a later stage only; modellers have shied away from
publishing their model as open source out of fear that having to add
good practices is too demanding. We here argue for making a habit of
following a list of simple and not so simple practices early on in the
implementation of the model life cycle. We contextualise cherry-picked
and hands-on practices for supporting Good Modelling Practice, and we
demonstrate their application in the example context of the Viable North
Sea fisheries socio-ecological systems model.
\end{abstract}




\date{\today}
\end{frontmatter}
\begin{tikzpicture}[remember picture,overlay,shift={(current page.north east)}]
\node[anchor=north east,xshift=-4cm,yshift=-1cm]{\href{https://www.openmodelingfoundation.org}{\includegraphics[width=3cm]{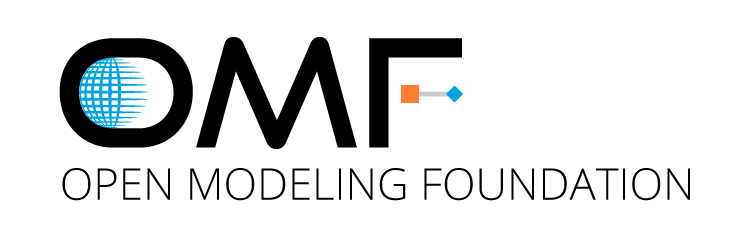}}};
\node[anchor=north east,xshift=-1cm,yshift=-1cm]{\href{https://www.hereon.de}{\includegraphics[width=3cm]{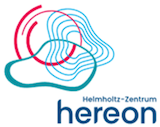}}};
\end{tikzpicture}

\section{Introduction}\label{introduction}

Frequently in socio-environmental sciences, models are used as tools to
represent, understand, project and predict the behaviour of these
complex systems, and for many more purposes (Epstein 2008; Edmonds et
al. 2019). The degree of a model's formalisation ranges from conceptual
to mathematical equations to implementation in software (Romanowska
2015), and--by definition--all of these models are purpose-driven
simplifications of the system they represent (Stachowiak 1973). We here
concentrate on computational models, i.e.~on socio-environmental models
implemented in software, and there are many of those out there:
Currently the CoMSES Net (2024) lists 1153 agent-based models (ABM, see
list of abbreviations provided); two decades ago, there were 1360
ecological models (Benz, Hoch, and Legović 2001); in 2015, a survey
amongst 42 modellers counted 278 aquatic ecosystem models (A. B. G.
Janssen et al. 2015).

So computational models are plenty and omnipresent in our field and they
have proven their value in being ``fruitfully wrong'' (Epstein 2008).
They often, however, escape strict falsifiability: the code may be
verifiable only within certain accuracy ranges; the model may be
validated only in site-specific application (Refsgaard and Henriksen
2004). The more useful assessment for its scientificity is that of
fitness-for-purpose, given its purpose, scope and assumptions (Edmonds
et al. 2019; Hamilton et al. 2022; Wang et al. 2023). Good Modelling
Practices (GMP) aim at ensuring this.

Examples of such practices often named are: a clear purpose, a thorough
domain understanding, going from simple to complex, ensuring
reproducibility, exploring sensitivities and validation with good
quality data (e.g., Crout et al. 2008). The first reference to GMP may
have been by Smagorinsky (1982), who claimed that ``under any
circumstance, good modelling practice demands an awareness of the
sensitivity {[}{]} to parametrization'' (p.16). From here on, GMP were
elaborated and became widespread in the field of hydrology with its
first handbook on the topic (Van Waveren, R et al. 1999); it has since
been applied to all areas of socio-environmental sciences and has been
adopted in community standards such as the Overview, Design, Detail
(ODD) documentation protocol and its derivatives (Grimm et al. 2006,
2010, 2020).

GMP appear as five phases in the model life cycle (Jakeman et al. 2024,
their Figure 1): (1) Problem scoping, (2) conceptualisation, (3)
formulation and evaluation, (4) application, and (5) perpetuation, and
the reiteration of these phases. Good Modelling Software Practices
(GMSP) are prominent in phases 3--5 and are thus subsumed under GMP. But
where GMP is concerned with the reliability and validity of a model,
GMSP is concerned with the quality and reliability of a model's software
implementation and beyond: a good quality model software is not
restricted to good computer code, but will support iterations through
the model life cycle and will enable Good Scientific Practice (DFG 2013;
All European Academies 2023).

Other scientific and technical fields where software plays a major role
developed the concept of Good Software Practice (GSP); it can be traced
back to the origins of the Unix system, which has at its core not a
monolithic but highly granular structure of many little programs that
``do one thing only, and do it well'' (Ritchie and Thompson 1974). These
little tools should allow to develop new software in a better way,
argued B. W. Kernighan and Plauger (1976), for example by following the
DRY principle: ``Do not repeat yourself'', and to KISS --- ``keep it
simple, stupid!'' (Johnson 1960; Raymond 2003).

Cotemporaneously the Free Software movement emerged to emancipate
software from the ownership of companies and consider it a public good,
to be (1) run for any purpose, (2) studied and modified, (3)
distributed, and (4) modified and distributed (Stallman 1983, 1996), and
with it the practices to ensure these freedoms in OSS. Educating about
GSP became central in projects such as the Software Carpentry (Wilson
2016), highlighting the utility of live coding, pair programming and
``open everything''. Plenty of checklists for following GSP have been
published (e.g., Artaza et al. 2016; Wilson et al. 2014).

For Open Data, internationally agreed criteria are that they are
Findable, Accessible, Interoperable, and Reusable (Wilkinson et al.
2016). Consequently, also the software evaluating these data should be
FAIR: (F) easy to find; (A) retrievable via standardised protocols; (I)
interoperable with other software by data exchange or an application
programming interfaces (API), and (R) understandable, modifiable, and to
be built upon, or incorporated into other software (Barker et al. 2022).
Research organisations like CoMSES Net educate about FAIR research
software, including rich metadata, code management, archiving,
documentation, and licensing\footnote{\url{https://tobefair.org} CoMSES
  Net initiative to make models FAIR}. Fairness should include
recognition of the work of the people that develop such modelling
software but are often not acknowledged in publications as the major
form of scientific output--the research software engineers (RSE, Katz et
al. 2019; Hettrick et al. 2022).

Despite many existing GMP and GSP guidelines, however, much of the model
source code corpus---roughly 80\% (Barton et al. 2022)---is not
published at all along with its scientific publication, and for a
trivial reason: Barnes (2010)'s survey states that ``the code is a
little raw'' was named as the main reason for not publishing the model.
Here we aim to address this fear and help build confidence that the
model code is good enough (Wilson 2016; Barnes 2010). Beyond those, we
break principles down to concrete tools that implement good practices
during the modelling software creation process.

\phantomsection\label{fig:triangle}
\begin{figure}
\centering
\pandocbounded{\includegraphics[keepaspectratio]{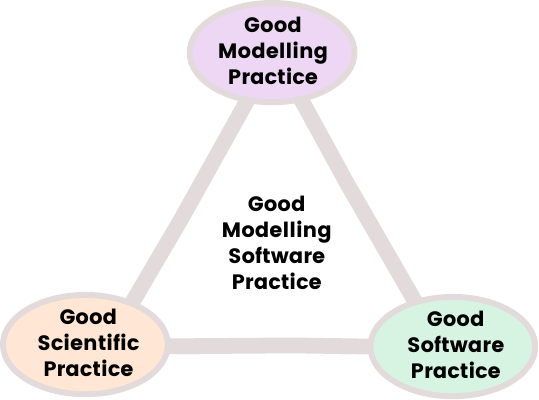}}
\caption{The triangle formed by good practices in modelling, software
and research \label{fig:triangle}}
\end{figure}

We start off by motivating each of the good practices and
contextualizing them towards the goal of publishing a model software or
a scientific result arising from it; we put GMSP at the center of a
triangle formed by good practices in modelling, software development,
and research (\hyperref[fig:triangle]{Figure 1}); we categorize broadly
which practices must, which ones should and which ones may be
implemented, summarized in \hyperref[fig:blocks]{Figure 2}. We describe
the tools that can be used in a non-exhaustive way covering the entire
range of good practices; you may and will deviate from the tools we
selected, or disagree with them, and you may add others or leave some
out that we suggested.

\section{Good practices for whom?}\label{good-practices-for-whom}

All models start with a purpose, and the purpose must be stated (Edmonds
et al. 2019), that has since long been the number one GMP. It does not
hurt to know about your domain, either; speak to other experts, develop
a conceptual model, only then start formalizing your model in math and
in software (Wang et al. 2023; Romanowska 2015; Grimm et al. 2006) to
arrive at your computational model---a purpose-driven and simplified
representation of your system in software. This model is by and for
yourself, it is for readers and reviewers, and it is by and for your
collaborators, constituting three tiers of user groups that are targeted
by tools addressing GMSP.

\subsection{Single authors -- yourself}\label{single-authors-yourself}

Socio-environmental modelling software can be created by a \emph{single
person}; in fact, it often is in student or postdoc projects or
individual programming sprints (Hettrick et al. 2022). Unfortunately,
from a software development perspective, the necessity of having to be a
domain expert at the same time means that 90\% of scientists are
self-taught developers without formal training in software engineering
(Wilson 2016). So that one person will not only have to write and read
her own code and to apply it for simulations, she will have to align the
software development with her research, will have to understand what she
did in the past, she will have to retrace the steps that led her to the
current state. All of this necessitates to some degree that the work is
stored redundantly in backups, and that changes are documented. She
would like to ensure that code changes do not break previous work and
does that by testing the model after every update (Rosero, Gómez, and
Rodríguez 2016).

Authors of scientific models eventually become authors or co-authors of
scientific publications arising from or with the help of a model. As
scientific authors, they are bound by ethical considerations such as the
Guidelines for Safeguarding Good Scientific Practice, among them the
unambiguous declaration intellectual property (IP) to the work and
ensuring availability of the model for a prolonged period of time (DFG
2013; All European Academies 2023). The IP declaration carries with it
the proper acknowledgement of software the model is built on, and
respecting the copyrights and permissions pertaining to third-party
sources used. The archiving requirement carries with it the obligation
to ensure that technical failures or changes in the circumstances of the
model author do not lead to the loss of the model: decentralised backups
or public repositories help.

\subsection{Reviewers and readers}\label{reviewers-and-readers}

If the model is to be used to produce scientific results subject to
\emph{peer} review, the single person will have to ensure
reproducibility of results. She will have to subject it to an editor or
a reviewer, thus make it readable and understandable, and document it.
And sometimes, she might be asked to support the reviewers in executing
the model somewhere outside her own computer infrastructure. When
feedback comes, there should be a platform to file the individual
concerns and address them.

To be published in a scientific journal, reviewers need to be able to
access and understand the model. Under the hood, the write-good-code
advice is old and simple: ``Write code that minimises the time it would
take someone else to understand it---even if that someone else is you''
(Flesch 1950). But it also needs to be easy for a peer to execute your
model and to understand its inputs and outputs. Various documentation
principles can help to ensure this accessibility, including
automatically generated application programming interface (API)
references, in-code documentation and the generation of metadata through
packaging.

But even before a reviewer invests her time in evaluating a model
software, much can be done by the author herself. In fact, where a
reviewer might concentrate on model validation, the author could ensure
model verification (Sargent 2010). Verification answers the questions:
Does the code do what the author intends it to do? To ensure this, it
has become GSP to employ unit testing and try to reach a good coverage
of the test framework over your entire code base (Ellims, Bridges, and
Ince 2006), and to automate verification with the source code managment
(SCM) service as Continuous Integration (CI, Shahin, Ali Babar, and Zhu
2017).

How does an author deal with the feedback she receives during a friendly
or journal-led review? Often, this comes as an itemised list of points
to address; as such it is in an ideal form to be converted to
``tickets'' or ``issues'' in the SCM service, or a dedicated issue
tracker system linked to the model software. Improvements to the model
code can then be tied to the issue tracker, transparently documenting
the resolution of those issues, and helping to formulate the rebuttal to
reviewer critique.

\subsection{Collaborators}\label{collaborators}

If at least one \emph{other} person is using the model, the
permissions---also known as the license terms---become pertinent. This
other user needs a way to communicate with the developer, for feature
requests or for reporting bugs. If that person intents to improve on
your work, the permissions become more important, needing contributor
agreements and codes of conduct. How are decisions made about the now
joint modelling software?---some form of governance model needs to be
established. With the growth of a community, even a community management
system might be required, with granular access, distributed roles, and
fine-grained permissions.

Regardless of the size of the collaboration, structured reviews,
pre-commit hooks, and common coding standards can be used to maintain
high code quality. Software needs to be sustainable as you aim for
establishing a larger user base, use it in other scientific software,
but most importantly, use the model to support scientific conclusions:
the model code is considered primary data and must be available a decade
after a scientific publication relying on it (DFG 2013); but will your
model be re-runnable and reproducible for that period of time (Benureau
and Rougier 2018)? Hardware and software environments change, as might
the developer's focus of work, all contributing to code rot (Liew 2017)
or software collapse (Hinsen 2019). Project funding, short-term
contracts, and mobility requirements stimulate frequent staff or job
rotations. Releasing a model as Free and Open Source Software (FOSS) may
be a way to facilitate sustaining a modeling software across those
rotations.

It is often desirable to collaboratively develop the software,
i.e.~involve another person or persons in improving the software. With
this collaboration come legal and governance decisions, as well as
technical requirements. The legal once concerns copyrights of the
different contributors, and often of the employers (research
organisations): Many academic institutions do not have yet clear
guidelines on the legal aspects of how to contribute to collaborative
software or how to accept contributions by other institutions. These can
be established in contributor agreements. Collaboration is also am
insurance against accidents: the ``truck factor'' asks: ``How many
people can get hit by a truck (or bus), before the project becomes
unmaintainable?''; the Open Source Security Foundation (2024) gold
standard requires that a project's truck factor is at least three.

For collaborative development with a smaller group, a governance system
known as the benevolent dictator is frequently encountered. A benevolent
dictator in software development refers to a leadership style where one
individual, often the project's creator or lead developer, has
significant control over decision-making processes within the project.
Despite holding considerable authority, this individual typically
exercises their power with the best interests of the project and its
community in mind, hence the term ``benevolent.'' This leadership model
aims to maintain direction and cohesion within the project while still
allowing for contributions and feedback from other team members or
contributors (Schneider 2022). In scientific projects, the governance
and licensing terms for a collaboration can be formulated as part of a
Memorandum of Understanding or consortium contract.

\section{Tools for Good Modelling Software
Practice}\label{tools-for-good-modelling-software-practice}

The tools described here can roughly be categorised as version control,
source code management system, licensing, documentation, packaging, good
code, archiving and maintenance, and publication.

\subsection{Version control software}\label{version-control-software}

A transparent and reproducible distributed SCM (formerly version control
system, VCS) is the basis for good software and has been termed
``possibly the biggest advance in software development technology''
(Spolsky 2010). The currently dominant SCM software is Git\footnote{\url{https://git-scm.com}
  OSS distributed VCS. Note: all Uniform Resource Locators (URL) in this
  manuscript have been last visited and checked on September 16, 2024},
originally invented by Linus Torvalds, the creator of Linux. As source
code is text, the SCM tracks changes in lines or parts of lines of text.
It can be very well used to manage other kinds of changing texts, such
as the text of this manuscript. In fact, this manuscript was started
with \texttt{git} \texttt{init;} \texttt{git}
\texttt{add\ manuscript.md;} \texttt{git} \texttt{commit} \texttt{-m}
\texttt{"feat:} \texttt{Created} \texttt{manuscript"}. Some care should
be taken with the latter commit messages, as they should be short and
descriptive to humans, and at best also machine-interpretable, e.g., by
following the conventional commits\footnote{\url{https://www.conventionalcommits.org}
  How to add human and machine readable commit meaning} recommendations.
Learning Git could be considered a valuable investment in yourself for
any creative work.

With graphical interfaces to Git, such as Sourcetree\footnote{\url{https://www.sourcetreeapp.com}
  Free Git client}, or various integrations in text editors, such as
Visual Studio Code\footnote{\url{https://code.visualstudio.com}
  Cross-platform editor with built-in Git}, it is now easy to visually
follow the step-wise development and provenance of code (or text
documents), go back to points in time or critical development stages, to
open experimental deviations from main development (\texttt{git}
\texttt{branch}) and combine diverging developments (\texttt{git}
\texttt{merge}). Did you mess up? Simply retrace your step back
\texttt{git} \texttt{revert}; it helps you even to find in the recorded
history those developments where things might have unnoticingly gone
wrong with \texttt{git} \texttt{bisect}.

Git and others are most powerful as distributed VCS, in combination with
other locations on your own computer, an intranet or the internet, for
saving your work in different places, the repositories, while keeping
all those versions synchronised. The interaction of two repositories is
managed by the unidirectional synchronisations \texttt{git}
\texttt{pull} and \texttt{git} \texttt{push}. These commands can be used
to synchronise the managed code also across different SCM services,
effectively allowing redundant and distributed backups and thus
minimizing the risk of losing the software from technical or human
errors or the risk of vendor lock-in (Nyman and Lindman 2013).

\begin{quote}
You \textbf{\emph{must}} have version control and you
\textbf{\emph{must}} have distributed redundancy.
\end{quote}

\subsection{Source code management
service}\label{source-code-management-service}

The most prominent online SCM service is GitHub\footnote{\url{https://github.com}
  Public GitHub SCM service}, but many academic institutions also offer
on-premise or cross-institutional dedicated SCM services, such as the
community GitLab of the German Helmholtz Association\footnote{\url{https://codebase.helmholtz.cloud}
  Community Gitlab of the German Helmholtz Association}, for their
students and researchers. A good reason to choose GitHub is the higher
number of potential contributors on this platform, estimated at roughly
100 million. Many automated tools for supporting software development
only work on GitHub. On the other hand, on-premise GitLabs may be
preferred by academic institutions, because the code is then hosted in
the research centre or by a dedicated subcontracted partner and may
offer better data and IP protection; it may also offer more elaborated
services or dedicated computing resources.

An SCM service is the entry point for collaborators to contribute,
provides a ticketing system and release management, and it offers
functionalities for CI and continuous deployment (CD, also known as
continuous delivery) of the software. Some SCM services also facilitate
project management workflows, such as milestone progress tracking; task
boards can be used for task tracking, work assignment or Agile
development (Beck et al. 2001).

\subsubsection{Ticketing system}\label{ticketing-system}

Often things do not work right away, or an error is detected in the
software. For this, SCM services offer ticketing (also called bug
tracker or issue tracking) systems, where one records the occurrence of
an error, a bug report, or a wish for future development, a feature
request. This works well for a single person, but even better when
collaborators and reviewers of the software record their observations on
faulty or missing issues with the software on this ticketing system. Git
commit messages can be linked to these issues mentioning the
hash-prefixed issue number, and they should carry a ``fix:'' or
``feat:'' prefix when they resolve a bug or implement a new
feature\footnote{\url{https://www.conventionalcommits.org} How to add
  human and machine readable commit meaning}. Beyond the ticketing
system, the SCM service may also offer communication facilities like
discussion forums, wikis, mailing list or service desks, which often
provide cross-referencing functionality to Git commits and issues.

\subsubsection{Continuous integration and
deployment}\label{continuous-integration-and-deployment}

CI is a development practice where developers integrate code into a
shared repository frequently, ideally several times a day, but at least
on every \texttt{git} \texttt{push} to a repository. Each integration is
then verified by an automated build and automated tests to detect errors
as quickly as possible---and to correct or recover from them (Shore
2004); the frequent and automated checks reduce the risk of accumulating
errors. SCM services like GitLab or GitHub provide such automated
integrations, but there are many CI tools available outside the
comprehensive SCM services, such as Circle CI\footnote{\url{https://circleci.com}
  Dedicated CI utitility}. CD is often triggered after the CI ends with
success. In this automation, the products of a modelling software can be
provisioned, such as a complete binary package for download, an updated
and nicely formatted documentation, or a suite of simulation results and
their statistical evaluation, for example. CD often interacts with other
external services to update web pages\footnote{\url{https://about.readthedocs.com}
  Popular documentation service ReadTheDocs}, to upload to package
repositories, or to submit to an archiving service.

\subsubsection{Pull requests}\label{pull-requests}

Pull requests (PR) offer an SCM service managed way to accept other
people's changes to your software into your development. Collaborators
typically duplicate your software in a fork, apply changes locally and
then file a PR providing a detailed explanation of the modifications and
their purpose. The SCM service allows you to review the changes,
possibly ask for further explanations or modifications, to have it
automatically tested with CI, and finally to \texttt{git} \texttt{merge}
the collaborator's work, and even to automate acknowledging the
contribution. Similarly, a branch-based approach to PRs uses separate
feature \texttt{git} \texttt{branch} branches and often allows to create
a Changelog\footnote{\url{https://keepachangelog.com} Keep a ChangeLog
  principles} based on the merging of the feature branch into main.

\begin{quote}
You \textbf{\emph{must}} use a collaboration platform, you
\textbf{\emph{should}} use CI, you \textbf{\emph{may}} use CD.
\end{quote}

\subsection{Licensing}\label{licensing}

Model software development is a creative process. It thus constitutes IP
and the right to determine how the model software can be used, re-used
and shared and modified, i.e.~the copyrights. The exact terms are laid
down in what is called a license. Without a license there is no
permission, so every model software needs a license, and with it the
name of the person or organisation holding the copyrights. While some
model software may be published under proprietary licenses and without
disclosing the source code, the majority of current modelling software
is distributed as open source software (OSS\footnote{\url{https://opensource.org/osd}
  The Open Source Definition by the Open Source Initiative 2024}), and
under a permissive or copyleft license, among them the widely used BSD,
MIT, Apache and GPL licenses.

There are strategic decisions involved in choosing for copyleft versus
permissive licenses, also related to the community in your field and
dependent on third-party software used in your modelling software paper.
There are tools to support choosing a license\footnote{\url{https://choosealicense.com}
  Choose an OSS license}, to manage licenses towards better
reuse\footnote{\url{https://reuse.software} Verification for OSS
  licensing}, and to assess the compatibility of different licenses with
a project\footnote{\url{https://www.fossology.org} Checks for OSS
  license compliance}\textsuperscript{,}\footnote{\url{https://github.com/oss-review-toolkit/ort}
  OSS Review Toolkit for managing (license) dependencies}. Some research
software are dual-licensed, to provide at the same time an OSS license
for the research community and the public, and a proprietary one for
commercial use.

\subsubsection{Contributions}\label{contributions}

With collaboration also comes the obligation to sort out the copyrights
evolving from different contributors, who are all creators and thus
natural copyright holders (or their organisation). Your contributors may
choose to assign their copyrights to you in what is usually called a
copyright transfer agreement (CTA) and is well known from the
publication process for scientific papers before the Open Access (OA)
movement. Alternatively, your contributors may permit you to exercise
copyrights arising from their contribution in a separate agreement, a
Contributor License Agreement (CLA) or a Fiduciary License Agreement.
Project Harmony\footnote{\url{https://www.harmonyagreements.org} Project
  Harmony contributor agreements} or the Contributor
Agreements\footnote{\url{https://contributoragreements.org} Contributor
  Agreements infrastructure} support the drafting of such agreements.

\subsubsection{Collaboration}\label{collaboration}

Collaborative software engineering is an intensely people-oriented
activity (Singer, Sim, and Lethbridge 2008); to keep both the software
as well as the collaboration healthy, many projects adhere to the
Contributor Covenant\footnote{\url{https://www.contributor-covenant.org}
  Contributor Covenant codex}. This provides guidelines for respectful
and constructive behaviour, it prohibits harassment and discrimination,
and overall help maintain a positive and welcoming environment.

Traditionally in science, mostly the authors of scientific publications
are acknowledged and cited. But for scientific software, there are more
roles that do not qualify for authorship: foremost those of the research
software engineers, but also, e.g., data managers, administrators,
friendly reviewers, high-performance computing (HPC) support staff. The
All Contributors\footnote{\url{https://allcontributors.org} Contributor
  acknowledgement automation} bot integrates with the GitHub SCM service
and facilitates proper acknowledgement of those and other non-author
groups.

\begin{quote}
You \textbf{\emph{must}} have a license for yourself and contributions
and you \textbf{\emph{should}} have guidelines for collaboration.
\end{quote}

\subsection{Versioning and Releasing}\label{versioning-and-releasing}

In contrast to standard journal publications, software is always a
work-in-progress (WIP). Even if there are no new features implemented,
software needs to receive regular updates because of the rapid
technological development -- both in hardware as well as software.
Without updates to the software, scientific analysis may become
irreproducible because the software cannot be used (Liew 2017; Hinsen
2019); but also, heterogeneous and scattered incremental updates of the
software may lead to ``code decay'', where the global code quality
declines despite local improvements (Eick et al. 2001).

This need for continuous maintenance requires tracking the state of
software and recording changes to software beyond the capabilities
offered by \texttt{git} \texttt{log}. In an SCM, each state can be
marked with an identifier \texttt{git} \texttt{tag} that carries a human
readable short description as well as version information following a
consistent strategy, such as semantic\footnote{\url{https://semver.org}
  Semantic versioning} or calendar versioning\footnote{\url{https://calver.org}
  Calendar versioning}. Releases are an elaborated version of tags, and
can include further resources, such as additional documentation or
pre-compiled binaries. They also integrate with archiving services such
as Zenodo upon a new release.

While changes to the source code are tracked in the SCM, the reasoning
behind those and the user-focused communication of these changes should
be kept in a change log\footnote{\url{https://keepachangelog.com} Keep a
  ChangeLog principles}, a technology since long enforced by the GNU
coding standard (Chen et al. 2004).

\begin{quote}
You \textbf{\emph{should}} have versioning.
\end{quote}

\subsection{Bundling your application}\label{bundling-your-application}

Much of research software, and especially those (like climate models)
with a long history, are distributed by giving access to the repository
hosted on an online SCM service, or to an archive file that contains the
entire source code. Users are then expected to install the (often
numerous) requirements, and eventually compile the code and run the
model. To improve usability, more state-of-the art software engineering
techniques like packaging and containerisation are available (Vanegas
Ferro et al. 2022). Not only do they standardise and simplify the
installation of the code, but they also improve its findability via
machine- and human-readable metadata, support reusability via versioning
and ensure proper citeability. We refer to this as bundling your
application. This bundling can be integrated in the CD by deploying the
bundle in the associated registries of the online SCM
services\footnote{\url{https://docs.github.com/en/packages}}\textsuperscript{,}\footnote{\url{https://docs.gitlab.com/ee/user/packages}}.
You can distribute a package of your model software or your model
software including a runtime environment

\subsubsection{Packages}\label{packages}

Packages are commonly used in programming languages to standardise and
simplify the installation of software, and to make the software findable
via machine- and human-readable metadata. Packages are files that
contain other files, most importantly a manifest that tells the package
name and version and lists its content. There are language-specific
package managers, e.g., for Python, Julia, R, NPM or Fortran, and
language-independent package managers, such as Debian's \texttt{dpkg} or
Continuum's \texttt{conda}\footnote{\url{https://conda-forge.org} Conda
  Forge community packages}; the latter, however, often depend on the
operating system and computer architecture. Packages provide a detailed
declaration of dependencies, including version constraints. Tools like
\texttt{versioneer}\footnote{\url{https://github.com/python-versioneer/python-versioneer}
  Versioning automation} provide the possibility to combine the package
version with Git tags. Even the smallest analysis scripts can be
distributed in form of a package, while some large models can be hard or
impossible to package.

Packages can be distributed via package registries to increase
visibility and availability of the software (Allen 2019). The metadata
contained in the package is consumed by the registries and made
available as a catalogue, such as the Python Package Index\footnote{\url{https://pypi.org}
  Prominent python package registry}.

\subsubsection{Container images}\label{container-images}

Installation can further be simplified for the user by distributing the
software as a container image. Like packages, container images are
distributed as a single file. They contain an entire operating system
layer where the necessary software and its dependencies have been
installed already. These images can then be published in container
registries to make them reusable. A prominent tool to build containers
is Docker\footnote{\url{https://www.docker.com} Docker tool to build
  containers}, a prominent registry the Docker Hub\footnote{\url{https://hub.docker.com}
  Free to use container image registry}, or the built-in container
registries of the online SCM services GitHub and
GitLab\footnote{\url{https://docs.github.com/en/packages}}\textsuperscript{,}\footnote{\url{https://docs.gitlab.com/ee/user/packages}}.
Apptainer\footnote{\url{https://apptainer.org} HPC container builder
  (formerly Singularity)} provides container imaging for for High
Performance Computing (HPC).

Distributing a model as a container makes it independent of the user's
infrastructure and thus more easily reusable and reproducible (Vanegas
Ferro et al. 2022).

\begin{quote}
You \textbf{\emph{may}} provide packages or containers.
\end{quote}

\subsection{Documentation}\label{documentation}

Proper documentation often determines a model software's impact and
usability. Most developers use inline comments for code, and they are
indeed important, but their impact on helping other scientists in
contributing or using the software is limited. Above all, proper Readme
text files (often in Markdown format as \texttt{ReadMe.md}) are
important, even if the author is its sole user. The Readme is usually
the first file added to a new project; it is also the first entry point
for users of the software to understand what he or she is looking at, as
SCM services render it prominently on your project's start page as a
hypertext (HTML) representation. A Readme provides an overview of the
modeling software: its purpose, assumptions and scope in narrative form.
A well-crafted Readme enhances user experience by providing clear,
concise, and relevant information on using the model. It can also
include a brief description of the software's architecture and its main
components, as well as installation instructions. It should contain a
section about how you want your software to be cited, or where to find
the citation instructions (Allen 2019). A helpful template covering
these sections is available at makeareadme\footnote{\url{https://www.makeareadme.com}
  \texttt{ReadMe.md} template}.

A contributing guide fosters a collaborative environment and encourages
contributions from other researchers. It often appears as a
\texttt{Contributing.md} file or as a section in the Readme; it provides
guidelines on how to contribute to the software, the coding standards to
follow, and the process for submitting changes. This ensures that the
software continues to evolve and improve, benefitting the entire
research community.

Tools like Sphinx\footnote{\url{https://www.sphinx-doc.org} Sphinx
  documentation tool} or MkDocs\footnote{\url{https://www.mkdocs.org}
  YAML-based documentation tool} automate the process of building
documentation and can contribute to creating user-friendly model
software more with more ease: By including inline code comments, they
support an up-to-date API documentation; additionally, they support
\emph{doctests}, which ensures that the examples in the documentation
work as expected, enhancing the reliability of the documentation.

Important sections in any documentation are:

\begin{enumerate}
\def\labelenumi{\arabic{enumi}.}
\item
  Detailed and clear \emph{installation instructions}, that eliminate
  guesswork, making the software accessible to a broader audience. These
  instructions should cover various operating systems and potential
  issues that might arise during the installation process. They should
  also specify the prerequisites, such as required libraries or
  dependencies.
\item
  A \emph{user manual} (in its minimal form a \emph{getting started
  guide}) is essential for enabling other modellers to rerun and use
  your code. It provides step-by-step guidance on how to use the
  software, explains the functionality of different modules, and shows
  examples of their use. This allows users to evaluate the model's
  fitness for purpose and apply it effectively.
\item
  \emph{API documentation} provides a detailed description of how the
  software's functions work, the parameters and data they accept, and
  the output they return. This is crucial for users who want to
  integrate the software into their own code or use it for more complex
  tasks---or reuse parts of your model in their modelling software.
\item
  If you aim for contributions by other researchers, the \emph{developer
  manual} is a must-have for the onboarding. It contains more detailed
  information about the framework that you are developing, that do not
  have place in the contributing guide or user manual.
\end{enumerate}

\subsubsection{Self-checks and badges}\label{self-checks-and-badges}

Part of the documentation may also be devoted to promotion,
self-checking, and community building. For these, software badges have
become widespread. They are little visual indicators put atop your
Readme that inform readers and yourself at first glance about diverse
aspects of your model and modelling process (Lee 2018). Things to show
are the activity of your development, the status of passing the CI, the
percentage of code covered by tests; information about portability and
software security, the status of self-assessments, or the publication
status and DOI, amongst many others. Badge awarding can be formally
categorised and reviewed (NISO 2021), but they can also be created by
yourself\footnote{\url{https:/shields.io} Badge creation}; motivating
yourself to keep to good practices may be the most important factor in
badge awarding, and showing badges has been shown to positively
correlate with data sharing (Kidwell et al. 2016). Particularly devoted
to an entire array of Good Practices is the Open Source Security
Foundation's OpenSSF badge program\footnote{\url{https://www.bestpractices.dev}
  OpenSSF badges}. It consists of many self-assessment questions, and at
the end reports the practices in your project at a basic, silver, or
gold level. Of course, you should show off this badge in your project's
Readme!

\begin{quote}
You \textbf{\emph{must}} have documentation in a Readme, you
\textbf{\emph{should}} elaborate documentation.
\end{quote}

\subsection{Clean and correct code}\label{clean-and-correct-code}

Software code must be readable fast not only by a computer but also by
humans (Flesch 1950). Code and community conventions are key here; e.g.,
the PEP8 conventions for Python ensure consistency across different
softwares and facilitate automated automated code formatting and
linting.

\subsubsection{Code formatting and
linting}\label{code-formatting-and-linting}

Automated formatters help adhere to conventions. Tools like
Black\footnote{\url{https://black.readthedocs.io} Python formatter} or
isort\footnote{\url{https://pycqa.github.io/isort} Python import sorter}
adjust (Python) code to meet specific formatting guidelines, eliminating
the need for manual formatting and ensuring consistency across the
codebase. A general-purpose tool for many file formats and languages is
Prettier\footnote{\url{https://prettier.io} General purpose code
  prettifier}.

Language-specific linters, such as ESLint\footnote{\url{https://eslint.org}
  Javascript linter}, flake8 or linter-formatters such as
Ruff\footnote{\url{https://docs.astral.sh/ruff} Python linter and
  formatter}, go a step further by running also checks against coding
quality rules\footnote{\url{https://www.flake8rules.com} List of Python
  coding standards for automated verification}. They all provide
feedback that can help developers improve their software quality and
adhere to best practices.

Linters and formatters can be combined in pre-commit hooks\footnote{\url{https://pre-commit.com}
  Triggers checks before \texttt{git} \texttt{commit}}. Pre-commit hooks
facilitate the formatting and linting process by automatically running a
list of tools before each \texttt{git\ commit}. This ensures that all
committed code adheres to the defined conventions and standards.

\subsubsection{Code Structure}\label{code-structure}

Good code is probably the oldest of all good practices, going back to
the single-purpose Unix tools (Brian W. Kernighan and Pike 1999). This
philosophy has found its way into larger programs, which are broken down
into manageable and readable building blocks, often functions or
modules, that should have a clear and documented way to interact with
data and the rest of the code, the API. Even if your code is not a
library for use by others, an internally consistent API can help you to
encapsulate functionalities, to allow data manipulation only via
dedicated accessors, to be explicit about acceptable inputs, and to
finally write safer and more correct code.

Separate concerns in the code structure: do not mix functionality with
visual appearance, avoid mixing input and output with processing. Try to
implement error handling, then try to break your own code. Often, there
is no need to code a functionality yourself, better re-use someone
else's working solution from an existing library or piece of code, ``be
ruthless about eliminating duplication'' (Wilson et al. 2017): vice
versa, all parts of your modelling software that have an API may also be
more easily reusable by others. Together with human-readable and
consistent naming practice, which may adhere to community conventions,
following this advice should provide simplicity, generality, and
clarity, the ``bedrock{[}s{]} of good software'' (Brian W. Kernighan and
Pike 1999).

\subsubsection{Code verification}\label{code-verification}

As code gets more complex, it is important to be able to ensure that the
smaller units of the code function as expected. For this, the concept of
unit tests has been developed, where in principle every code unit has a
mirrored part either within the code or separate from it in a testing
framework. So, every time you write a new function, think about (and
implement) how its functionality can be tested. How much of a code is
covered with test is called the coverage, and the coverage can help you
identify parts of the code that are not well tested. But even if all
units of a code are correct, their interaction may not be, and a model
might be dysfunctional or produce unreasonable results. To catch these,
regression and reproducibility tests (RT) should be regularly run
(Rosero, Gómez, and Rodríguez 2016).

\begin{quote}
You \textbf{\emph{must}} provide readable code and you
\textbf{\emph{should}} use automated tools for formatting, linting, and
verification.
\end{quote}

\subsection{Archiving}\label{archiving}

Archiving the source code of the model and post-processing routines is
essential Good Scientific Practice (DFG 2013). The simplest solution for
open-source projects is using the Software Heritage Project\footnote{\url{https://www.softwareheritage.org}
  Software archive crawler}. Many SCM services are already crawled by
this project, so when you upload your public repository, it will be
archived automatically; you can also add your project manually.

A more citeable and permanent archiving solution is guaranteed by
uploading the source code and generating a Digital Object Identifier
(DOI). There are various institutional platforms that offer this
service, the most prominent one is likely Zenodo\footnote{\url{https://zenodo.org}
  Zenodo archive}. Most of these platforms additionally offer to create
versioned DOIs, where each upload gets its own DOI, and there is one
single DOI that represents all versions of the model software. Used in
combination with releases, each release gets its own DOI and the
contributions to the release are citeable separately. Helpers exist to
automate this upload, such as Zenodo's GitHub integration\footnote{\url{https://docs.github.com/en/repositories/archiving-a-github-repository/referencing-and-citing-content}
  Zenodo integration}, or the hermes workflow\footnote{\url{https://github.com/hermes-hmc/hermes}
  Hermes workflow} (Druskat et al. 2022).

\begin{quote}
You \textbf{\emph{must}} archive and you \textbf{\emph{should}} do this
on a community platform.
\end{quote}

\subsection{Publish your model}\label{publish-your-model}

Many scientific journals nowadays request to publish the source code
that has been used to generate figures alongside the journal
publication. This increases usability and tractability, it does not,
however, acknowledge the amount of work that has been put into the
software or its documentation (Hettrick 2024); nor does it take into
account that the software might be developed further after the paper has
been published.

Consequently, software may be published separate from the scientific
publication in a dedicated software journal, often allowing to track the
development across multiple versions. Such software journals, e.g.~the
Journal of Open Research Software or the Journal of OSS (JOSS, Smith et
al. 2018) are meant to be developer-friendly, i.e.~they integrate with
the development workflow and tools used to follow GSP.

Such software journal publications are not domain-specific, so they do
not generally improve the findability of your software. Therefore, the
most important aspect on publishing your code (or its metadata) is to
list it in a platform that your colleagues know, such as the communities
on Zenodo\footnote{\url{https://zenodo.org/communities} Zenodo
  communities}. An alternative might be the Research Software
Directory\footnote{\url{https://research-software-directory.org}
  Helmholtz Research Software directory} (Spaaks et al. 2020) that also
allows to group software into projects and interests; or, the community
targeting ABMs on the CoMSES Net, where code, metadata and documentation
is optionally reviewed (M. A. Janssen et al. 2008). An overview of
software registries is provided by SciCodes\footnote{\url{https://scicodes.net/participants}}.

\begin{quote}
You \textbf{\emph{must}} publish your model software.
\end{quote}

\subsection{Software development
templates}\label{software-development-templates}

The multitude and the complexity of GSP techniques and tools put a
burden on the modeller. Not only does it require acquiring new
knowledge, the rapid technological development makes it also difficult
to stay up to date. The diverse skill sets of researchers and time
constraints further complicates the situation (Wilson 2016).

To jumpstart new modelling software projects GMSP, they can be derived
from templates that already provision these techniques from the start of
a project (Pirogov 2024). Cruft\footnote{\url{https://cruft.github.io/cruft}
  Cruft template manager} and cookietemple\footnote{\url{https://github.com/cookiejar/cookietemple}
  Template creation} keep things up to date behind the scenes. Sommer,
Saß, and Benninghoff (2024), e.g., provide a fork- and cruft-based
methodology for modelling ecosystems: they provide a standardised setup
for post-processing routines, plugins, etc., implementing many of the
techniques and tools mentioned in this manuscript.

\begin{quote}
You \textbf{\emph{may}} use templates.
\end{quote}

\section{Good enough modelling software practice - a use
case}\label{good-enough-modelling-software-practice---a-use-case}

Viable North Sea (ViNoS) is a socio-ecological model of the German North
Sea coastal fisheries (Lemmen et al. 2023, 2024) coded in NetLogo
(Wilensky 1999) and embedded in a larger software system containing
data, and Python/R data pre- and postprocessing scripts. Its NetLogo
source code is broken down into a central NetLogo file and several
included object-specific module files. The code, data and scripts are
managed by Git, with a primary SCM service on an academic
Gitlab\footnote{\url{https://codebase.helmholtz.cloud/mussel/netlogo-northsea-species}}
and a secondary one on public Github. On the primary SCM, GitLab issues
provide the ticketing system and milestone planning. Upon each
\texttt{git} \texttt{push} to the repository, a CI rule triggers license
checking, builds Docker images for different versions of the NetLogo
IDE, performs unit and replicability testing within these containers.
Using them for subsequent CD, a small production simulation generates
model results and pushes them to a static web page on the SCM service;
the documents provided with the model (ODD, JOSS, this manuscript, and
others) are compiled from their version-controlled Markdown sources to
pdfs (using pandoc and LaTeX) and uploaded. The CD is integrated with
Mkdocs and pushes to a public readthedocs instance for the user guide.
On the secondary CI, a release management hooks into Zenodo to provide
permanent and DOI citable archives for each model release. Locally, a
pre-commit workflow, triggered upon each \texttt{git} \texttt{commit}
ensures that the copyrights and license catalogues are complete for each
file, that all structured documents comply with their respective type
definition, and that python codes conform to PEP8 coding standards. The
tools used for this are \texttt{reuse}, \texttt{black}, \texttt{flake8},
more general formatters to remove empty line endings, syntax checkers
for metadata in \texttt{yaml}, \texttt{toml}, and \texttt{json}
structured formats, and the general-purpose code beautifier
\texttt{prettier}.

In the program root folder, a \texttt{CITATION.cff} suggests how to cite
the software, a \texttt{codemeta.json} provides package meta
information, a \texttt{ChangeLog.md} the user-oriented change
information. Every directory contains at least a \texttt{ReadMe.md}. The
project implements the Contributor Covenant, its
\texttt{Contributing.md} file explains that the best way to contribute
is via issues on the SCM service and uses the Project Harmony template
for providing the legal infrastructure for the contributor agreements.
Badges shown are for CoMSES Net Open Code, Zenodo and JOSS DOIs, an
active repository status, the OpenSSF best practices and Contributor
Covenant adherence, reuse compliance, prettier code and code quality A.

For reproducibility, a conda \texttt{environment-dev.yaml} for
satisfying developer dependencies, as well as a \texttt{pyproject.toml}
for the python dependencies, are provided. The model has been archived
on Zenodo, and on CoMSES.net together with its ODD (Lemmen et al. 2023).
A software paper has been published in JOSS (Lemmen et al. 2024) with a
transparent review process\footnote{\url{https://github.com/openjournals/joss-reviews/issues/5731}
  Transparent review}; the model has been catalogued in the ABM specific
CoMSES.net as well as the Helmholtz research software directory.

\section{Conclusion}\label{conclusion}

Good Practices have been maturing in the areas of software, modelling,
and research. Adopting all of them for socio-environmental modelling
establishes Good Modelling Software Practices: Reproducibility and
citeability meet automated verification, rights management and
collaboration meet purpose-driven simplification and validation. For
this, you \textbf{must} have version control and distributed redundancy,
use a collaboration platform, have a license and documentation, provide
readable code, and archive and publish your model software. You
\textbf{should} use CI, have guidelines for collaboration, elaborate on
documentation, employ versioning and releases, use automated tools for
formatting, linting, and verification and target a community platform.
You \textbf{may} use CD and packages.

\phantomsection\label{fig:blocks}
\begin{figure}
\centering
\pandocbounded{\includegraphics[keepaspectratio]{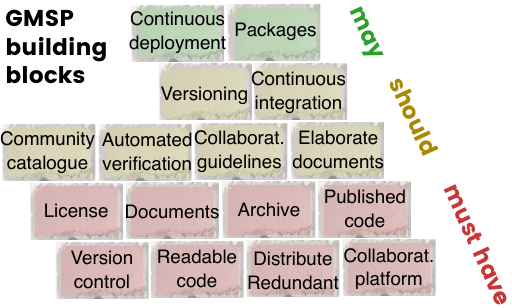}}
\caption{Must haves, should haves and may haves in Good Modelling
Software Practise \label{fig:blocks}}
\end{figure}

For many scientific model authors, getting stuff done may be more
important than documenting it thoroughly, usefulness is more valued than
red tape, spontaneous ideas are implemented preferably over those layed
down in a management plan; individual agency supersedes organisational
processes. Scientific model software development reflects the ideas
formulated in agile development: ``Individuals and interactions over
processes and tools. Working software over comprehensive documentation.
{[}{]} Responding to change over following a plan. And while the items
on the left are valued more, the items on the right are still
important'' (Beck et al. 2001). In this spirit, Good Modelling Software
Practice means that you do not have to do it all at once. The supporting
tools are numerous, and most follow a philosophy to do one thing only
and to do it well: they can be learned and employed one by one. Publish
your modelling software---it is good enough (Allen 2019; Barnes 2010)!
By learning and applying one tool at a time, everyone can acquire a
corpus of practices that eventually lead to better modelling
software---and better research (Katz et al. 2019).

The recent surge in Artificial Intelligence (AI) applications has
created an array of emerging good practices in the areas of security
(Polemi and Praça 2023), data treatment (Makarov et al. 2021), and in
supporting software development (Pudari and Ernst 2023), amongst others.
They will need to be negotiated by society but eventually also find
their place in supporting Good Modelling Software Practices.

\section*{References}\label{references}
\addcontentsline{toc}{section}{References}

\phantomsection\label{refs}

\vspace{6pt}

\textbf{Acknowledgements: }This research is funded by the programme
Changing Coasts of the Helmholtz Gemeinschaft and an outcome of the
Multiple Stressors on North Sea Life (MuSSeL) project funded by the
German Ministry of Education and Research (BMBF, grant
\href{https://foerderportal.bund.de/foekat/jsp/SucheAction.do?actionMode=view&fkz=03F0740A}{03F0862A}),
and the DataHub Project of the Helmholtz Association's Changing Earth
programme. This manuscript benefitted from discussions in the context of
the Open Modeling Foundation (OMF,
\url{https://www.openmodelingfoundation.org}). We thank the OSS
communities that make ours and other modellers' work possible, among
them the developers of and contributors to Linux, Git, Python, NetLogo,
pandoc, and LaTeX.

\textbf{Author contributions: }C.L.: Conceptualisation, Methodology,
Visualisation, Software, Writing -- original draft, Writing -- review \&
editing. P.S.: Writing - original draft, Writing - review \& editing.


\subsection*{The following abbreviations are used in this manuscript}\label{sec:abbreviations}

\noindent
\begin{tabular}{@{}lp{0.8\columnwidth}}
ABM & Agent-based Model \\
AI & Artificial Intelligence \\
BSD & Berkeley Software Distribution \\
CalVer & Calendar versioning \\
CC & Creative Commons \\
CD & Continuous Deployment \\
CFF & Citation File Format \\
CI & Continuous Integration \\
CLA & Contributor License Agreement \\
CoMSES & Network for Computational Modeling in the Social and Ecological
Sciences \\
CTA & Copyright Transfer Agreement \\
DOI & Digital Object Identifier \\
DRY & Do not repeat yourself \\
FLA & Fiduciary License Agreement \\
FOSS & Free and Open Source Software \\
GMP & Good Modelling Practice \\
GMSP & Good Modelling Software Practice \\
GNU & GNU is Not Unix \\
GPL & GNU General Public License \\
HTML & Hyper Text Markup Language \\
IP & Intellectual Property \\
JOSS & Journal of Open Source Software \\
JSON & JavaScript Object Notation \\
KISS & Keep it simple, stupid! \\
MIT & Massachusetts Institute of Technology \\
MuSSeL & Multiple Stressors on North Sea Life \\
OpenSSF & Open Source Security Foundation \\
OA & Open Access \\
ODD & Overview, Design, Details \\
OMF & Open Modeling Foundation \\
OOP & Object-oriented programming \\
OSS & Open Source Software \\
OSI & Open Source Initiative \\
PEP & Python Enhancement Proposal \\
PR & Pull Request \\
RSE & Research Software Engineer \\
RT & Regression Testing \\
SCM & Source Code Management \\
SemVer & Semantic versioning \\
SSI & Software Sustainability Institute \\
URL & Uniform Resource Locator \\
VCS & Version Control System \\
ViNoS & Viable North Sea \\
WIP & Work in progress \\
YAML & YAML Ain't Markup Language \\
\end{tabular}

\end{document}